\documentclass[twocolumn]{aastex631}

\shorttitle{Helium Nova V445 Pup Is Not a SN{\rm I}a Progenitor}
\shortauthors{Schaefer}
\graphicspath{{./}{figures/}}

\begin{document}
\title{The Unique Helium Nova V445 Puppis Ejected $\gg$0.001 M$_{\odot}$ in the Year 2000 and Will Not Become a Type {\rm I}a Supernova}

\author[0000-0002-2659-8763]{Bradley E. Schaefer}
\affiliation{Department of Physics and Astronomy,
Louisiana State University,
Baton Rouge, LA 70803, USA}

\begin{abstract}

V445 Puppis is the only known example of a helium nova, where a layer of helium-rich gas accretes onto the surface of a white dwarf in a cataclysmic variable, with runaway helium burning making for the nova event.  Speculatively, helium nova can provide one path to produce a Type {\rm I}a supernova (SN{\rm I}a), within the larger framework of single-degenerate models.  Relatively little has been known about V445 Pup, with this work reporting the discovery of the orbital period near 1.87 days.  The companion star is 2.65$\pm$0.35 R$_{\odot}$ in radius as an evolved giant star stripped of its outer hydrogen envelope.  The orbital period immediately before the 2000 eruption was $P_{\rm pre}$=1.871843$\pm$0.000014 days, with a steady period change of (-0.17$\pm$0.06)$\times$10$^{-8}$ from 1896--1995.  The period immediately after the nova eruption was $P_{\rm post}$=1.873593$\pm$0.000034 days, with a $\dot{P}$ of ($-$4.7$\pm$0.5)$\times$10$^{-8}$.  The fractional orbital period change ($\Delta P/P$) is $+$935$\pm$27 ppm.  This restricts the mass of the gases ejected in the nova eruption to be $\gg$0.001M$_{\odot}$, and much greater than the mass accreted to trigger the nova.  So the white dwarf is losing mass over each eruption cycle, and will not become a SN{\rm I}a.  Further, for V445 Pup and helium novae in general, I collect observations from 136 normal SN{\rm I}a, for which any giant or sub-giant companion star would have been detected, yet zero companions are found.  This is an independent proof that V445 Pup and helium novae are not SN{\rm I}a progenitors.

\end{abstract}

\section{Introduction}

With insight, the existence, nature, and properties of `helium novae' were predicted in 1989 by Kato, Saio, \& Hachisu (1989).  The idea is that a white dwarf (WD) in a cataclysmic variable (CV) might accrete helium-rich gas, which accumulates until a helium shell flash powers a normal-looking nova eruption.  The accumulated helium might come from accretion off a giant star stripped of its outer hydrogen-rich envelope, or from a degenerate star made mostly of helium (like for an AM CVn star).  The ejecta of a helium nova should be essentially hydrogen-free.  

Helium novae might be a separate path to create a Type {\rm I}a supernova  (SN{\rm I}a), see Kato \& Hachisu (2003) and Kato et al. (2008).  One of the premier open questions in modern astrophysics concerns the nature of the progenitors of normal SN{\rm I}a systems (the Progenitor Problem), see Maoz, Filippo, \& Nelemans (2014) for a review.  The progenitors certainly are a close binary star with one member being a CO WD.  One popular solution to the Progenitor Problem is that the companion star of the CO WD is a second WD, hence there are two degenerate stars in the binary for the so-called double-degenerate (DD) model.  The other popular solution is to have the companion being a normal non-degenerate star, hence the binary has only one degenerate star for a so-called single-degenerate (SD) model.  

V445 Pup (Nova Puppis 2000) was discovered in eruption on 30 December 2000 by K. Kanatsu (Kanatsu 2000).  Prediscovery images show the star to be in quiescence up until 26 September 2000, and at 8.8 mag (nearly the discovery magnitude) on 23 November (Ashok \& Banerjee 2003).  The peak magnitude might be near 8.8 or possibly substantially brighter.   The eruption spectra were startling for the utter lack of any hydrogen lines, and this is the key point for identifying V445 Pup as a helium nova (Kato \& Hachisu 2003).  Otherwise, the strong emission lines of various metals resembles that of slow novae.  The light curve faded from peak by 3 magnitudes ($t_3$) in 240 days, like for slow novae.  In February 2004 at McDonald Observatory, while seeking a photometric time series to discover an orbital period, I was startled to find that V445 Pup appeared $\sim$6 mags fainter in the $R$ band than before the nova eruption.  V445 Pup had gone into a dust-dip (like for DQ Her and for D-class novae) caused by dust formation in the dense ejecta.  Startlingly, this dust dip was uniquely and extraordinarily deep (getting to 6 mag fainter than pre-eruption) and long lasting (it has only recovered to $V$=17 even 24 years later).  So the ejected shell must be very massive.  Startlingly, Woudt et al. (2009) found a fast-expanding finely-structured symmetrical bipolar nebula, expanding at speeds up to 8450 km s$^{-1}$.  This places the Earth close to the equatorial dust disk, which causes the deep dust dip.  In all, we have a unique and extreme nova, with V445 Pup being the only known example of a helium nova.

Other than the recognition as a helium nova, little is known about the fundamental nature of V445 Pup.  The orbital period is not known\footnote{Goranskij et al. (2010) used old plates from Moscow and Sonneberg to claim a periodicity of 0.650654 days.  But this is just one peak amongst 15 roughly-equal alias peaks ranging from 0.39 to 3.74 days, with plotted folded light curves for most.  Their chosen alias was selected because they liked the Lafler-Kinman periodogram and they saw an insignificant dip that they took to be an eclipse.}.  Modelers (Kato et al. 2008, Shen \& Bildsten 2009, Brooks et al. 2015) expect orbital periods from roughly one hour (for AM CVn systems) to several days (for stripped giant companions). The distance is poorly determined.   The spectral energy distribution (SED) of the pre-eruption star has been variously fit to that of an A0 {\rm V} star (Goranskij et al. 2010) or an accretion disk (Ashok \& Banerjee 2003), but this was constructed with magnitudes from 5 widely separated times, such that the plot is disjointed and can only be a bad fit to any blackbody or power law.  Rather, the only useable information is that the SED does not suffer any large turnover in the $B$-band, so the stellar temperature of the companion must be $>$10,000 K.  The mass of the dust formed in the ejecta has been estimated from the infrared flux (Ashok \& Banerjee 2003, Lynch et al. 2004, Banerjee et al. 2023), but any such estimate has uncertainties of $>$3 orders-of-magnitude due to the usual uncertainties in dust temperature, distance, as well as the dust composition and size distribution.  Further, to go from any such badly-known measure of the dust mass to the mass of the ejecta would require a dust fraction that is uncertain by orders of magnitude.  Goranskij et al. (2010) were even suggesting that the original WD had been completely destroyed by the nova eruption.

As the only known example of a helium nova, V445 Pup has long excited theorists anxious to find an SD progenitor (e.g., McCully et al. 2014, Kool et al. 2023).  But there has been no useful observational advance since 2010.  Kato et al. (2008) gives their final conclusions as ``We emphasize the importance of making observations after the dense dust shell disappears, especially observations of the color and magnitude, orbital period, and inclination angle of the orbit. These are important to specify the nature of the companion.''.   Woudt et al. (2009) gives ``Validation of the white dwarf+helium star model as the appropriate binary configuration of V445 Puppis will come from the determination of its orbital period.''

\section{Orbital Period}

The orbital period $P$ is expected to be in the range from one hour up to weeks.  To find the period as a coherent optical modulation, I had previously made an ineffective photometric time series at McDonald Observatory and I had previously looked through a short series of Harvard and Sonneberg sky photographs (plates), again with no significant periodicity.  With the completion of the DASCH program\footnote{The Digital Access to a Sky Century @ Harvard (DASCH, J. Grindlay P. I.) program has recently completed the ambitious and important digitization of the entire archival collection of sky photographs from 1887 to 1989 at the Harvard College Observatory (HCO).  The entire archive and light curves are now freely available at \url{http://dasch.rc.fas.harvard.edu/lightcurve.php}.}, additional plates and magnitudes became available, so I constructed a combined light curve, with 175 magnitudes from 1896 to 1989.  A discrete Fourier transform then immediately gave a single peak that is highly significant, a period near 1.87188 days.  This period has no ambiguities or problems with any aliases or artifacts of the window function.  A fold on this period (see Fig. 1) shows a typical CV orbital modulation with a broad minimum of total amplitude 0.40 mag.  This is obviously the orbital period.

\begin{figure}
	\includegraphics[width=1.01\columnwidth]{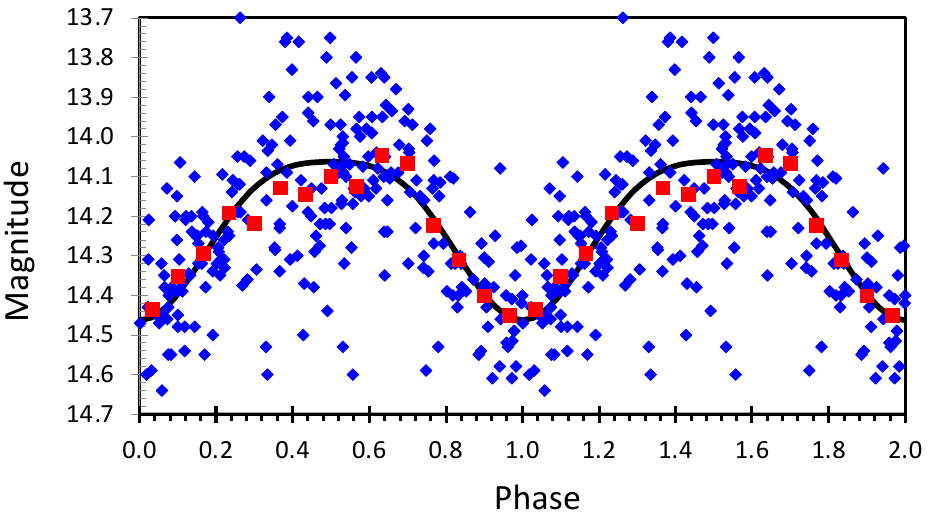}
    \caption{Pre-eruption light curve for one full century.  This folded light curve is made from 264 archival plates 1896--1995 (see Table 1), with the individual $B$ magnitudes (small blue diamonds) having a typical photometric uncertainty of $\pm$0.10 mag.  The phase-binned light curves are shown with the red squares.  The black curve shows a smoothed template for the orbital variations.  We see a typical CV light curve with a broad minimum and a flat maximum.  A primary point of this plot is that the pre-eruption (before 2000) light curve has a well-defined period and it is greatly different from the post-eruption orbital period (see Fig. 2).} 
\end{figure}

Now, in hindsight, the exact same periodicity is seen independently in both the 51 Moscow plates from 1969--1989 and the 32 Sonneberg plates from 1984--1991.  Goranskij et al. (2010) had actually had the correct period as his most-significant peak, but had passed it over for a less-significant alias that displayed an insignificant dip that was taken to be an eclipse.  The best period from the Moscow plates is 1.87186 days.  

For the Sonneberg plates, back in 2013, I independently measured all the same plates as used by Goranskij, so I have simply averaged these to come up with the best Sonneberg light curve.  For the Sonneberg data alone, the periodicity is apparent as a marginally significant peak at 1.8714 days.  So I have found a reliable pre-eruption orbital period ($P_{\rm pre}$) that appeared consistently throughout three independent data sets 1896--1991. 

After the 2000 eruption was over, the dust dip kept the star faint enough that no one effectively could get a time series.  Then in 2018, the Zwicky Transient Factory (ZTF) started a long series of isolated measures in the $zg$ and $zr$ bands, for 929 magnitudes spread over six observing seasons from 2018 to 2024.  These two colors were individually detrended to remove the effect of the brightening star as it recovers from the dust dip.  The ZTF light curve was all taken from one longitude on Earth and only during the normal observing seasons, so we expect and see many daily and yearly aliases in any discrete Fourier transform.  For V445 Pup, the discrete Fourier transform has the highest power being for a peak close to $P_{\rm pre}$.  The detrended ZTF light curve shows a highly-significant peak at 1.8726 days for a time centered on 2020.447.  The real uncertainty on this period is difficult to calculate, mainly because the light curve is sampled with isolated points, with flickering and trends dominating over the periodic signal with full-amplitude of 0.08 mag.

All of my photometry from Harvard, Moscow, Sonneberg, ZTF and more are presented in Table 1.  $TESS$ fluxes are not included because their number is large and unwieldy, while being easily available\footnote{\url{https://mast.stsci.edu/portal/Mashup/Clients/Mast/Portal.html}} at the Mikulski Archives for Space Telescopes (MAST).

\begin{table}
	\centering
	\caption{V445 Pup Photometry ((full table with 1337 magnitudes in machine readable formation in the electronic article)}
	\begin{tabular}{lllll}
		\hline
		Julian Date   &   Year   &   Band   &  Magnitude  &  Source$^a$ \\
		\hline
2413880.8132	&	1896.88	&	{\it B}	&	14.01	$\pm$	0.10	&	HCO	\\
2414186.8811	&	1897.72	&	{\it B}	&	14.12	$\pm$	0.10	&	HCO	\\
2414259.8090	&	1897.92	&	{\it B}	&	14.10	$\pm$	0.10	&	HCO	\\
2414930.8610	&	1899.75	&	{\it B}	&	13.93	$\pm$	0.10	&	HCO	\\
2415403.6503	&	1901.05	&	{\it B}	&	14.05	$\pm$	0.10	&	HCO	\\
...	&		&		&		&		\\
2460341.8529	&	2024.08	&	{\it zg}	&	17.62	$\pm$	0.03	&	ZTF	\\
2460352.7763	&	2024.11	&	{\it zg}	&	17.69	$\pm$	0.03	&	ZTF	\\
2460352.8201	&	2024.11	&	{\it zr}	&	16.87	$\pm$	0.02	&	ZTF	\\
2460354.7562	&	2024.12	&	{\it zr}	&	16.85	$\pm$	0.02	&	ZTF	\\
2460368.7569	&	2024.16	&	{\it zr}	&	17.04	$\pm$	0.02	&	ZTF	\\
		\hline
	\end{tabular}	
	
\begin{flushleft}	
\
$^a$ Sources: {\bf HCO~}These $B$ magnitudes are from averages of my multiple by-eye measures from the Harvard plates and the DASCH measures from the same plates.  These magnitudes are in the modern $B$ magnitude system because the spectral sensitivity of the blue plates is the same as for Johnson's system and because the APASS comparison stars were used.
{\bf Sky Survey~}These magnitudes are measured from the all-sky surveys with the big Schmidt telescopes at Palomar, ESO, and the UK Schmidt, as calibrated from APASS comparison stars, and as reported in Goranskij et al. (2010).
{\bf Moscow~}These $B$ magnitudes were taken from the archival plates of the Moscow SAI Crimean Station 40-cm astrograph and calibrated with APASS comparison stars (Goranskij et al. 2010).
{\bf Sonneberg~}These $B$ magnitudes are from the archival plates at Sonneberg Observatory, and always calibrated with the APASS comparison stars.  The quoted magnitudes are averages of the values measured by Goranskij et al. (2010) and my own by-eye measures of the same plates.  Many of the plates are intentional double exposures, for which I made independent magnitude estimates of both exposures.
{\bf Goranskij~}These CCD magnitudes in BVRI were calibrated with APASS comparison stars and reported in Goranskij et al. (2010).
{\bf Monard~}Berto Monard (Bronberg Observatory, Pretoria, South Africa) made many CCD measures, all collected into the AAVSO International Database, see \url{https://www.aavso.org/data-download}.
{\bf APASS~}The AAVSO Photometric All-Sky Survey covers the entire sky to 17th magnitude with well-calibrated $B$, $V$, $g'$, $r'$, and $i'$ magnitudes, see \url{https://www.aavso.org/download-apass-data}.
{\bf ZTF~}The Zwicky Transient Factory magnitudes are in the $zg$ and $zr$ bands, with the light curve available at \url{https://irsa.ipac.caltech.edu/cgi-bin/Gator/nph-scan?submit=Select&projshort=ZTF}.
\\
\end{flushleft}	
	
\end{table}

The $TESS$ spacecraft has visited V445 Pup during three of its time intervals of 22-25 days each with nearly gap-free light curves, with Sector 7 centered on 2019.057, Sector 34 centered on 2021.077 (see Fig. 2), and Sector 61 centered on 2023.087.  The time resolutions of these light curves are 1800, 600, and 200 seconds, respectively.  Fourier transforms show highly-significant peaks at periods of 1.8850, 1.8682, and 1.8722 days respectively.  These variations are expected and normal due to the continuous flickering of the CV, with only 12-14 cycles per Sector.  The folded light curves (see Fig. 2) show a low amplitude modulation with a broad minimum and a relatively flat-topped maximum.  The observed $TESS$ amplitude is artificially small, because the $TESS$ pixels of 41 arc-seconds contain substantial additional light past just that of the target star.

\begin{figure}
	\includegraphics[width=1.01\columnwidth]{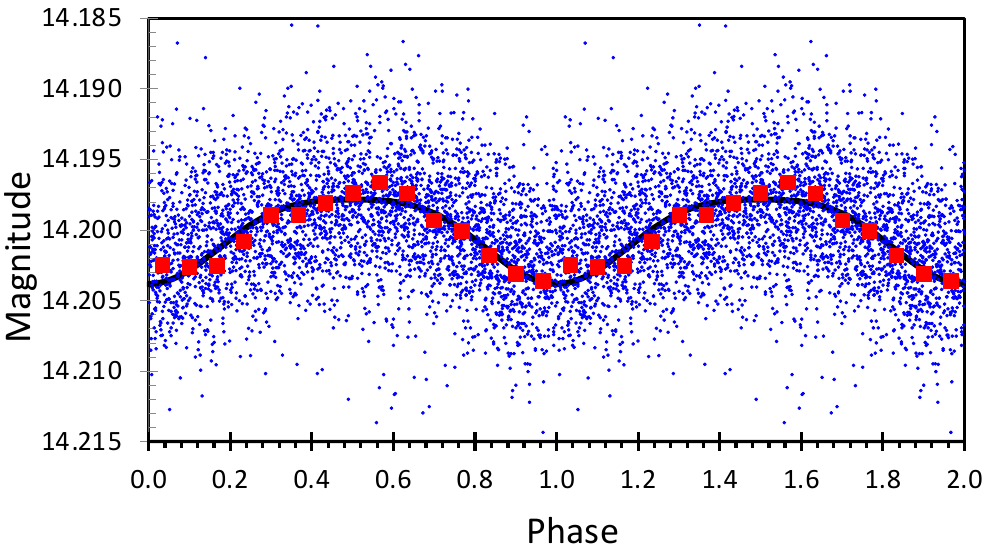}
    \caption{{\it TESS} light curve for Sector 34.  This folded light curve is made from 3430 fluxes (converted to magnitudes) from early 2021, as given by the small blue diamonds.  The typical error bar on each point is $\pm$0.003 mag, so much of the scatter is from ordinary Poisson noise.  The phase-binned light curve is shown with the red squares.  The same template as in Figure 1, except with scaling, is shown with the black curve.  The amplitude shown here is artificially low, because the {\it TESS} pixels contain extra light.  The shape is the same as before the 2000 eruption (see Fig. 1) and the same as a typical CV light curve.} 
\end{figure}

The Fourier transform of the combined ZTF and $TESS$ light curves (covering 2018--2024) shows a highly significant and unambiguous peak with a post-eruption orbital period of $P_{\rm post}$=1.8731$\pm$0.0005 days.  This is greatly different from the pre-eruption period.

The orbital period changes from a combination of a steady period change ($\dot{P}$) during quiescence plus a sudden period change ($\Delta P$=$P_{\rm post}$-$P_{\rm pre}$) across the 2000 eruption.  In terms of the $O-C$ curve, the shape is a broken parabola, where the pre-eruption $\dot{P}_{\rm pre}$ can be different from the post-eruption $\dot{P}_{\rm post}$.  The $O-C$ curve must be continuous across the eruption (because the stars cannot jump ahead or behind in their orbit), although its slope can appear to have a sudden kink upward (for a fast period increase across the eruption) or a sudden kink downward (for a fast period decrease across the eruption).  Indeed, the size of this kink yields the desired measure of $\Delta P$.  The times of minimum light can be modeled as
\begin{eqnarray}
T_{\rm min} = E_0 +NP_{\rm post} +0.5P\dot{P}_{\rm post}  N^2  ~~~~~~(N \ge 0), \nonumber \\
T_{\rm min} = E_0 +NP_{\rm pre} +0.5P\dot{P}_{\rm pre} N^2  ~~~~~~(N < 0).
\end{eqnarray}
With this choice, the $\dot{P}$ quantity is dimensionless, or it can be viewed as having dimensions of s/s.  I have chosen the epoch $E_0$ to be close to the start of the eruption soon after JD 2451850.  $N$ is always an integer, representing a sequential count of the times of minima.  The orbital phases for any light curve point can be calculated by subtracting off the time for the immediately previous $T_{\rm min}$, then divided by $P$.  The phase ranges  from 0.0 to 1.0, although Figs 1 and 2 plot duplicates for the range 1.0 to 2.0.

The best period model comes from a chi-square fit of all the light curve points to a template light curve (like the black curves in Figs 1 and 2) where the model predicted magnitude is based on the calculated phase.   This method has the strength that all the data are used, with allowance for the various photometric uncertainties.  The use of a chi-square allows a simple and reliable means of calculating the one-sigma error bars on the period changes.  The use of Equation 1 forces the before and after fits to be compatible, plus it allows the cycle count to be accurate.  And the use of a $\dot{P}$ term allows for the inevitable steady period changes.

With this, I have a comprehensive global fit involving 13,476 magnitudes from 1896--2024.  I find $P_{\rm pre}$=1.871843$\pm$0.000014 days and  $\dot{P}_{\rm pre}$=($-$0.17$\pm$0.06)$\times$10$^{-8}$.  The JD epoch is $E_0$=2,451,850.400$\pm$0.065.  For after the eruption, $P_{\rm post}$=1.873593$\pm$0.000034 days and  $\dot{P}_{\rm post}$=($-$4.7$\pm$0.5)$\times$10$^{-8}$.  With this, $\Delta P$=$+$0.00175$\pm$0.00005 days, or $\Delta P$$/P$=$+$935$\pm$27 parts per million (ppm).

The $\Delta P$ is huge, at a 2.5 minute period change.  This is by far the largest of the 14 $\Delta P$ measures\footnote{This is from Schaefer (2023), plus my recent measure of the 2020 U Sco eruption.}.  The one exception is for T CrB, with a 270 minute period change (Schaefer 2023), with this iconic red giant system involving some different period change mechanism.  The $\Delta P$ is positive, meaning that the period {\it increased} across the eruption, and that the binary separated substantially.  Further, the effects of mass loss from the ejecta dominate over the various mechanisms for angular momentum loss.  This large and positive $\Delta P$ forces a large mass for the nova ejecta.

The $\dot{P}_{\rm post}$ measure for V445 Pup is over a factor of two times larger in magnitude than for all other 14 novae with measures, again with the exception of T CrB (Schaefer 2023).  The measured V445 Pup $\dot{P}_{\rm pre}$ and $\dot{P}_{\rm post}$ values are negative, meaning that the quiescent helium nova has a steady period {\it decrease} over time.  The steady period changes in quiescence are a competition between the effects of the steady mass transfer (which must be a period increase for cataclysmic variables with the companion having less mass than the white dwarf) and of the effects of angular momentum loss by the binary (which must be a period decrease).  V445 Pup has steadily decreasing period in quiescence, so the angular momentum loss must be dominating.  Unfortunately, we have no observational evidence as to the mass accretion rate either before or after the eruption, due to the companion's luminosity hiding the accretion disk.  Unfortunately, for cataclysmic variables in general, the nature of the steady period change is not known\footnote{The mechanism is certainly not `magnetic braking' for a radiative giant helium star, nor for CVs in general, see Schaefer (2023, 2024).  The nature of this unknown mechanism for angular momentum loss in CVs is now the most important problem for all CV research.}, and the case of V445 Pup is just as unknown.  Whatever the nature of the angular momentum loss, V445 Pup has a much larger rate in magnitude than for all other regular novae, other than T CrB.  The next-largest values are all from novae with sub-giant companions, while all of 48 cataclysmic variables of all types with main sequence companions are all smaller than the sub-giants (Schaefer 2024).  I expect that the V445 Pup binary angular momentum loss rate does not depend on the helium composition of the companion or the accreted layer, so I do not see any reason for V445 Pup to be different from all other novae, in terms of $\dot{P}$.  But the nova systems with the largest measured $\dot{P}$ just happen to be the only two binaries with giant companion stars.  This suggests that the angular momentum loss in cataclysmic variables is connected to the luminosity class of the companion, where giant companions have the largest $|\dot{P}|$, main sequence companions have the smallest $|\dot{P}|$, and sub-giant companions are in the middle (Schaefer 2024).  I interpret this as hopefully providing a critical clue as to the nature of the angular momentum loss in cataclysmic variables in general.

\section{Companion Star Properties}

The nature of the V445 Pup companion star is critical for understanding its evolution.  The companion must certainly have a predominantly-helium composition in its outer layers, with this material falling onto the WD and then being ejected in the nova eruption.  The extreme cases range from a degenerate helium WD as in an AM CVn binary, up to a giant companion star that has been stripped of its outer hydrogen-rich layers.  These possibilities can be easily distinguished from the companion star radius as determined from the orbital period.

My newly-discovered $P$ is 1.87 days, which immediately confirms the primary aspect of the model by Kato et al. (2008).  From Kepler's Law and presumed stellar masses, we can get the size of the companion's Roche lobe (Frank et al. 2002), and the companion must just fill its Roche lobe.  

For this, we need the stellar masses.  Banerjee et al. (2023) ``speculate'' that $M_{\rm WD}$ is ``low'', as based on the low observed eruption amplitude, the long $t_3$, the large dust mass, and the low excitation spectrum.  However, the low amplitude is due to the high luminosity of the companion rather than any low-$M_{\rm WD}$.  The deductions from the other properties are dubious because they are using generalizations based on ordinary hydrogen-novae, while the case of helium-novae likely could be greatly different.   Contrariwise, the ejection velocity is up to 8450 km s$^{-1}$ (Woudt et al. 2009), and such is directly tied to the WD escape velocity, so $M_{\rm WD}$ must be maximal.  Kato et al. (2008) present a detailed physics model of helium novae, as applied to V445 Pup, deriving that $M_{\rm WD}$$\gtrsim$1.35 M$_{\odot}$, and I will adopt this value.  The mass of the companion star must be smaller than $M_{\rm WD}$ or so, as otherwise there would be runaway accretion.  For the companion to evolve off the main sequence in a useful time, the original mass must be larger than something like 1 $M_{\odot}$.  But the outer hydrogen-rich layers were stripped off, making for a greatly lower mass.

For $M_{\rm WD}$$\gtrsim$1.35 M$_{\odot}$ (Kato et al. 2008), and a companion mass ranging from 0.5 to 1.0 M$_{\odot}$, the companion's stellar radius $R$ ranges from 2.3 to 3.0 R$_{\odot}$, which can be represented as 2.65$\pm$0.35 R$_{\odot}$.  This immediately rejects the AM CVn possibility, and affirms the stripped-giant helium star possibility.  In particular, the early model of Kato et al. (2008) is confirmed.

For seeing the exact evolutionary status of the companion, we need to place it on the HR diagram, a plot of stellar luminosity ($L$) versus surface temperature ($T$ in degrees Kelvin) along with tracks of helium stars.  Just such a diagram appears as figure 6 of Kato et al. (2008), which I have used as the base for my Fig. 3.  The locus of a star of fixed radius is a straight line on the HR diagram.  With the usual scaling from our Sun, 
\begin{equation}
L = (R/R_{\odot})^2 (T/5770)^4  L_{\odot}.
\end{equation}
For the minimal temperature from the SED of 10,000 K, the companion's $\log[L/L_{\odot}]$ ranges from 1.68 to 1.90.  For a temperature of $10^{4.6}$ K (40,000 K), $\log[L/L_{\odot}]$ ranges from 4.08 to 4.30.  This defines a line for possible positions for the companion star, shown in red in Fig. 3, with the line thickness describing the uncertainties.

We also have a constraint based on the absolute magnitude of the pre-eruption nova.  The $V$ magnitude was 13.7$\pm$0.2 steadily over the prior century\footnote{Both Goranskij et al. (2010) and my own $B$-band photometry shows a stable average of $B$=14.1 with the usual superposed flickering and orbital variations.  Goranskij has the only useable pre-eruption color information, because his measures from the original Palomar plates were calibrated with the modern APASS comparison stars and because only this pair of plates is simultaneous.  This gives $B-R$ of 0.88 mag.  For any sort of smooth SED, this makes $B-V$$\approx$0.4 and $V$=13.7$\pm$0.2 or so.}.  The poorly-constrained distance\footnote{This distance includes the prior from the 8200 pc expansion parallax of Woudt et al. (2009), along with the appropriate uncertainty for that method.  {\it Gaia} does not yet report any parallax in the third data release, although as V445 Pup brightens, we can expect a future parallax measure with a large fractional uncertainty.} is 6272$^{+2754}_{-1246}$ parsecs (Schaefer 2022).  The $E(B-V)$ is 0.51$\pm$0.10 (Iijima \& Nakanishi 2008, Banerjee et al. 2023).  Then, the peak absolute magnitude, $M_V$, is $-$1.9$\pm$0.7.  Such a luminous system cannot have a significant contribution from an accretion disk, so the light from the companion star is dominating.  This absolute magnitude must be converted to a bolometric magnitude with the bolometric correction.  The bolometric correction for a 10,000 K star is -0.27 (Flowers 1996), while a surface temperature of $10^{4.6}$ K (40,000 K) has a bolometric correction of -3.58 (Nieva 2013).  The bolometric magnitude can be converted to luminosity with the use of the value for our Sun as $+$4.81 mag for the $V$-band.  With this, $\log[L/L_{\odot}]$ is 2.8$\pm$0.3 for T=10,000 K, and is 4.1$\pm$0.3 for T=40,000 K.  This defines a locus in the HR diagram depicted in Fig. 3 as an orange line, with the thickness representing the uncertainty from the $M_V$ constraint.

\begin{figure}
	\includegraphics[width=1.01\columnwidth]{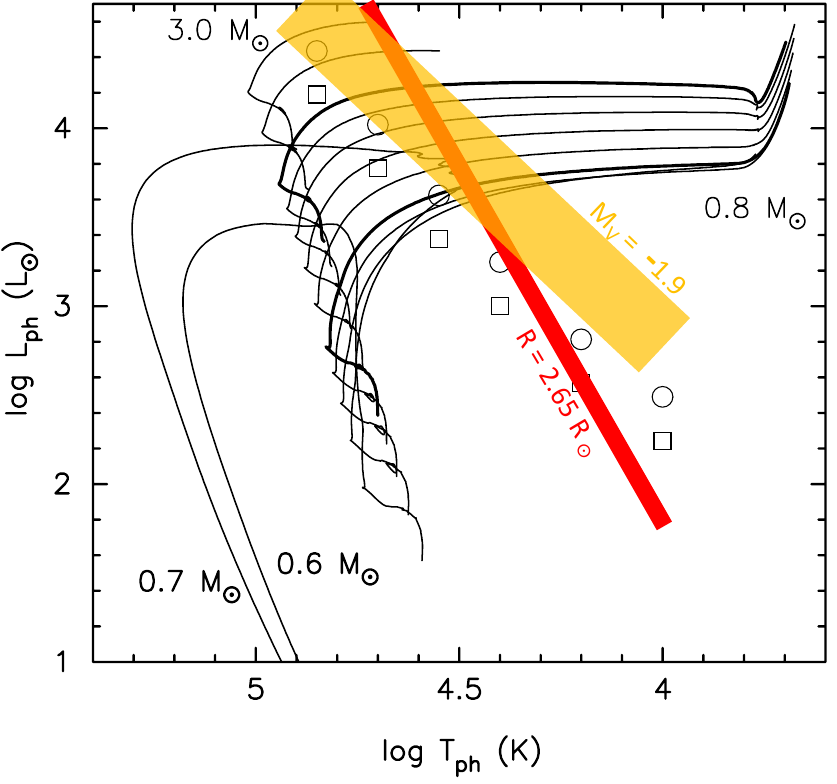}
    \caption{The V445 Pup companion star on the HR diagram.  The base diagram is from figure 6 of Kato et al. (2008), with the curves being tracks of helium star evolution.  The helium stars with original masses of 0.7 M$_{\odot}$ and less never get to the giant branch, while the stars with 2.5 M$_{\odot}$ or more have core carbon ignition at the point where the tracks stop.  The tracks that extend to the right are for original stellar masses of 0.8 to 2.0 M$_{\odot}$ with intervals of 0.2 M$_{\odot}$.  (The base diagram included the circles and squares as position of 14.5 mag stars at distances of 6.5 and 4.9 kpc, respectively, for visual extinction of 1.6 mag, as a depiction of the bolometric corrections.)  The constraint from the newly-discovered orbital period is that the stellar radius of the companion star is 2.65$\pm$0.35 R$_{\odot}$ is depicted as a thick red line, with the thickness representing the allowed region.  The constraint from the absolute magnitude ($-$1.9$\pm$0.7) with a bolometric correction is depicted with a thick orange line.  The companion star must lie in the intersection of the orange and red regions.  This constrains the companion to be a giant helium star recently evolved off the main sequence, just as calculated for the Kato et al. (2008) model. } 
\end{figure}

The two constraints (from $R$=2.65 R$_{\odot}$ and $M_V$=$-$1.9) cross in the HR diagram at close to 35,000 K with a luminosity of $10^{4.00\pm 0.11}$ L$_{\odot}$, or near to 10,000 L$_{\odot}$.  The joint constraint is a narrow slanted region in the HR diagram where the thick orange line covers the thick red line in Fig. 3.  The extreme range for the overlap of the two constraints stretches from 22,000 to 60,000 K and from 1800 to 59,000 L$_{\odot}$.

An additional constraint comes from the requirement that the current mass of the companion be less than $M_{\rm WD}$ or else the accretion would become runaway.  This cannot be readily depicted in Fig. 3, because we have no useful estimate of how much gas has been stripped off the companion.  So perhaps the original star was of 2 M$_{\odot}$, only to have lost one solar-mass of the outer solar-composition material stripped away to get to the current situation.  Further, the tracks are for those of original helium stars, so it is unclear where the tracks are for stars with their outer layers of hydrogen stripped away.

We now have a useful picture of the companion star, beyond the observation that the outer layers have no hydrogen.  The star is a giant, evolved off the main sequence, with a radius of 2.65$\pm$0.35 R$_{\odot}$, with $M_V$ equal to $-$1.9$\pm$0.7, with a luminosity of near 10,000 L$_{\odot}$, and a surface temperature of near 35,000 K.  The original mass of the helium star is unknown, but likely between 0.8 and 1.4 M$_{\odot}$.  This validates the specific model of V445 Pup from Kato et al. (2008), and rejects the various alternatives.

\section{Mass of the Ejecta For the 2000 Nova Eruption}

The mass of the gas ejected ($M_{\rm ejecta}$) by V445 Pup is critical for understanding the evolution.  The ejecta consists of the helium- and carbon-rich gases, plus the recently-formed dust.  No measure of the gas mass has been published, so we can only work with the measured dust mass .  Ashok \& Banerjee (2003) derive dust temperature of 1800 K and a dust mass of $10^{-9}$ M$_{\odot}$ for the early dust formation.  Lynch et al. (2004) use a dust temperature of 250 K and derive a dust mass of 1.5$\times$$10^{-5}$ M$_{\odot}$, for the corrected distance as in Woudt et al. (2009).  Shimamoto at al. (2017) modeled the infrared light with  cold (125 K) and warm (250 K) amorphous carbon components, for a total dust mass near 0.47$\times$$10^{-3}$ M$_{\odot}$, although they identify the recently ejected dust as the warm component for the dust mass of 0.018$\times$$10^{-3}$ M$_{\odot}$.  Banerjee at al. (2023) modeled the SED with  cold (105 K) and warm (255 K) components, for a total dust mass near 1.9$\times$$10^{-3}$ M$_{\odot}$, then assumed a gas-to-dust mass ratio of 10--100 to get $M_{\rm ejecta}$ of 0.01--0.1 M$_{\odot}$.  Banerjee further points out that the addition of the SEST 1.2 mm flux to the SED requires an additional cooler component at $\simeq$30--50 K and a dust mass of 0.01 M$_{\odot}$.  So the observational measures of the dust mass as reported in the literature vary by a factor of $10^7$.  Then, to get $M_{\rm ejecta}$, some unknown ratio of the dust-to-gas masses must be presumed.

Unfortunately, these $M_{\rm ejecta}$ estimates have a greatly larger real uncertainty than is expressed in the publications.  The reason is that the calculated estimates are highly sensitive to several parameters that are poorly measured or merely guessed at:  {\bf (A)~}The dust temperature is poorly measured.  We can see this from the wide range of reported temperatures in the previous paragraph.  And we can see this from the best of the SEDs incorporating infrared fluxes from 2003 up to 2012, such that the expected and observed large variations over time make for a confused SED of ill-determined temperature.  Ashok \& Banerjee (2003) show that the derived dust mass is proportional to the dust temperature raised to the $-$6 power.  With the order-of-magnitude variations in the published temperature estimates, the calculated dust mass will have real uncertainty of several orders-of-magnitude.  And indeed, the published dust masses range over many orders-of-magnitude.  {\bf (B)~}The dust fraction assumed by Banerjee has a range of one order-of-magnitude, yet this is based on experience from hydrogen-rich ejecta.  Rather, the dust fraction should be taken for hydrogen-poor and carbon-rich gas, which would likely be greatly different from Banerjee's assumption.  {\bf (C)~}The optical depth in the infrared is large (the dust dip is near 100$\times$ in the $K$-band), confounding the calculation that assumes the shell is optically thin.  {\bf (D)~}The dust mass scales as the square of the adopted distance.  The uncertainty in the distance used in the calculations is nearly a factor of 2$\times$, for an uncertainty in $M_{\rm ejecta}$ of 4$\times$, with this to be added in quadrature to the other uncertainties.  {\bf (E)~}Banerjee points out the possibility that substantial amounts of pre-existing dust might contribute to the measured SED, hence allowing the 2000 ejecta to contribute only a possibly-small fraction of the dust, with the uncertainty from this effects being unknown and possibly large.  With all these inconsistencies and problems, the calculated $M_{\rm ejecta}$ values have real uncertainties of $>$3 orders-of-magnitude for V445 Pup.

For novae in general, the question of $M_{\rm ejecta}$ has severe problems, both for theory and for observations.  For theory, various predicted values for the same nova eruption differ by up to two orders-of-magnitude (Schaefer 2011, Appendix A).  Observationally, the traditional measures from absolute fluxes from the hydrogen emission lines has six largely-irreducible sources of uncertainty, each of which has real error bars of one-to-three orders-of-magnitude (Schaefer 2011, Appendix A).  In the end, for both V445 Pup and novae in general, the estimates of $M_{\rm ejecta}$ have real uncertainties of several orders-of-magnitude.

There is one accurate and reliable method to measure $M_{\rm ejecta}$, but unfortunately this is a hard task operable for only rare novae, and this can only lead to a lower limit on the mass for a fraction of the cases.  The observational task is to measure the orbital period {\it before} the novae eruption ($P_{\rm pre}$), to measure the orbital period {\it after} the novae eruption ($P_{\rm post}$), and to use the change in the period ($\Delta P$=$P_{\rm post}$-$P_{\rm pre}$) to calculate $M_{\rm ejecta}$.  The physics comes from Kepler's Law, where the period depends on the total mass in the binary, so a change in the mass of the binary directly causes a change in the period of the binary.  The $\Delta P$ caused by the mass loss is 
\begin{equation}
\Delta P_{\rm ml} = 2  \frac{M_{\rm ejecta}}{M_{\rm comp}+M_{\rm WD}} P
\end{equation}
(see equation 6 of Schaefer 2020).  This equation has been derived in many papers over the decades, and it includes the loss of angular momentum to the binary as carried away by the ejecta, which is assumed to have the same specific angular momentum as the white dwarf.  The orbital periods are measured as a simple timing experiment of eclipses or photometric minima, so it is immune from the usual problems of distances, extinctions, dust fractions, filling factors, and ejecta temperatures.  The hard part is to get the series of {\it pre}-eruption timings, because no one was watching the star {\it before} the nova event, so we can only make do with archival sky images.  Few novae are bright enough to have adequate time coverage from archival data, and only a small fraction of those have an adequately large photometric modulation tied to the orbit.  As a career-long program started by me in 1990, I have measured $\Delta P$ across 12 nova eruptions of 10 nova systems (Schaefer 2023), while I have recently measured the fourth $\Delta P$ value for U Sco across its 2020 eruption.  Fortunately, V445 Pup is one of the rare cases where I can pull out an accurate and reliable $P_{\rm post}$ and $P_{\rm pre}$ (see Section 2).  So V445 Pup is my fourteenth $\Delta P$ measure.

The orbital period across a nova eruption will also change due to dynamical frictional angular momentum loss (FAML) as the companion star plows through the nova ejecta.  In the past, this $\Delta P_{\rm FAML}$ has been evaluated from a simplified case where the ejecta is all on a ballistic trajectory going out of the system, with this leading to an effect that is always smaller in magnitude than $\Delta P_{\rm ml}$, and of the opposite sign.  However, various groups (e.g., Shen \& Quataert 2022) have been realizing that such a calculation is missing the main effect caused by the companion star orbiting inside the outer edge of the hot envelope around the WD that is puffed up by the nuclear burning of the eruption.  This FAML is much larger in size (greatly more negative) than the old FAML calculation because the relative velocity between the companion and the accreting gas is smaller than for simple ejecta, and because the puffed-up envelope lasts greatly longer than a simple ballistic ejection.  The resultant $\Delta P_{\rm FAML}$ depends critically on the density and temporal structure of the envelope.  Unfortunately, the envelope properties cannot be measured and they are confused from contradictory theory models, so it is difficult to estimate the size of $\Delta P_{\rm FAML}$.  Crudely, we know that $\Delta P_{\rm FAML}$ must be small for fast novae with small $M_{\rm ejecta}$ (like for the recurrent novae) and must be large for slow novae with large $M_{\rm ejecta}$ (like for D- and J-class novae and for V445 Pup).  

The observed period change is just the sum of the effects from mass loss and FAML,
\begin{equation}
\Delta P = \Delta P_{\rm ml} + \Delta P_{\rm FAML}.
\end{equation}
Importantly, $\Delta P_{\rm ml}$ is always positive while $\Delta P_{\rm FAML}$ is always negative.  We generally cannot know $\Delta P_{\rm FAML}$ with much accuracy, but we at least know that $\Delta P_{\rm FAML}$$<$0 for all novae and $\Delta P_{\rm FAML}$$\ll$0 for slow novae with massive ejecta like V445 Pup.  So we can at least get a {\it limit} on  $\Delta P_{\rm ml}$, 
\begin{equation}
\Delta P_{\rm ml} \gg \Delta P,
\end{equation}
as applied for V445 Pup.  If the observed $\Delta P$ is negative, then this limit is not useful.  If the observed $\Delta P$ is positive, then we have a useful limit.  When combined with Equation 3, we have
\begin{equation}
M_{\rm ejecta} \gg 0.5 (M_{\rm comp}+M_{\rm WD})  \frac{\Delta P}{P}.
\end{equation}
So there we have it, an unambiguous and confident {\it limit} for the mass ejected for the case of V445 Pup.

For V445 Pup, the WD mass is $M_{\rm WD}$$\gtrsim$1.35 M$_{\odot}$ (Kato et al. 2008), while the companion star mass is 0.5--1.0 M$_{\odot}$ after the striping off of the hydrogen layer.  And $\Delta P$$/P$ is 935$\pm$27 ppm.  With this, $M_{\rm ejecta}$ is from $\gg$0.00087 to $\gg$0.00111 M$_{\odot}$.   So the ejecta mass is well represented as  $M_{\rm ejecta}$$\gg$0.001 M$_{\odot}$.

This limit on the ejecta mass is startlingly large.  The limit of $\gg$0.001 M$_{\odot}$ is one-to-four orders of magnitude larger than reported for any nova, or ever been theoretically predicted for any nova (e.g., Yaron 2005).  Presumably the difference is due to the helium-burning on V445 Pup rather than the hydrogen-burning on all other novae.

A main purpose of this paper is to test whether V445 Pup (and helium novae in general) is a SN{\rm I}a progenitor.  For this, I can test the SD requirement that the WD must be gaining mass over time.  This is a balance between how much mass is accreted between eruptions ($M_{\rm accreted}$) and how much mass is ejected from the system by each eruption ($M_{\rm ejecta}$).  SD models require $M_{\rm ejecta}$$<$$M_{\rm accreted}$.  Unlike for RNe, we do not have the systems recurrence time, nor do we even have any measure of the $\dot{M}$ either before or after the 2000 eruption.  Fortunately, the physics of thermonuclear runaways on a WD with accreted helium-rich gas is accurately known.  Detailed physics calculations can derive the accreted mass required to trigger the nova ($M_{\rm trigger}$).  And this trigger mass must equal the mass accreted between eruptions.  So the SD requirement is that $M_{\rm ejecta}$$<$$M_{\rm trigger}$.

Kato et al. (2008) provide the calculation of $M_{\rm trigger}$ for the conditions of V445 Pup.  For their models, $M_{\rm trigger}$ is always close to 0.00022 M$_{\odot}$.  They further calculate $\dot{M}$ of 4$\times$$10^{-8}$ M$_{\odot}$ yr$^{-1}$ and a nova recurrence time of nearly 5000 years.  

So the SD requirement is violated, because the accreted mass (0.00022 M$_{\odot}$) is not greater than the ejected mass ($\gg$0.001 M$_{\odot}$).  The violation is by a factor of $\gg$5$\times$.  That is, $M_{\rm ejecta}$ is much greater than $M_{\rm accreted}$, so the V445 Pup WD is {\it losing} mass over each eruption cycle, and the long-term evolution is for $M_{\rm WD}$ to be {\it decreasing} over time.  This is a forced consequence of my confident measure that the ejected mass is huge.  As such, V445 Pup is not a progenitor, and it can never become a Type Ia supernova.  

By extension from the only known example, helium novae are apparently not SN{\rm I}a progenitors.  However, we could imagine that V445 Pup is some sort of unusual extreme case with $M_{\rm ejecta}$$\gg$$M_{\rm trigger}$, while some of the other helium novae in our Galaxy have $M_{\rm ejecta}$$<$$M_{\rm trigger}$.  Given that our sample of one has the ejecta mass exceeding the criterion for `progenitorship' by over an order-of-magnitude, this possibility can only be of relatively low probability.  While there is no evidence for this possibility, it does mean that the progenitorship  is not absolutely denied for all helium novae.

\section{SN{\rm I}\lowercase{a} PROGENITORS DO NOT HAVE GIANT OR SUB-GIANT COMPANIONS}

I can make a second test of whether helium novae in general are SN{\rm I}a progenitors.  This test is to measure the fraction of normal SN{\rm I}a systems that have a helium giant companion.  If most of SN{\rm I}a systems have companions like in V445 Pup, then we have a solution to the global Progenitor Problem, and that is a single-degenerate solution.  If some not-small fraction of SN{\rm I}a systems have companions like in V445 Pup, then the Progenitor Problem has a partial solution from helium novae.  If the fraction of SN{\rm I}a systems with companions like in V445 Pup is zero, then this is proof that helium nova are not progenitors.

A variety of observational methods can be used to test whether any particular SN{\rm I}a has a companion like in V445 Pup.  One of these methods (seeking any ex-companion star) was pioneered by my group (Schaefer \& Pagnotta 2012, Edwards, Pagnotta, \& Schaefer 2012, Pagnotta \& Schaefer 2015), and this still has the deepest limit.  Other methods have been vigorously pursued by a number of other groups worldwide, with deep searches for the nearest supernovae and broad searches amongst many normal SN{\rm I}a.  These studies are summarized in Table 2.

\begin{table*}
	\centering
	\caption{Do SN{\rm I}a Progenitors have a Giant or Sub-giant Companion?}
	\begin{tabular}{llll}
		\hline
		Method   &   Supernova   &   Constraint   &  Reference$^a$ \\
		\hline
No visible progenitor	&	SN 2011fe in M101	&	$M_V >$ -1	&	[1]	\\
No visible progenitor	&	SN 2014J in M82	&	$M_V >$ -1	&	[2]	\\
No ex-companion	&	SNR 0509-67.5 in LMC	&	$M_V >$ +8.4	&	[3]	\\
No ex-companion	&	SNR 0519-69.0 in LMC	&	$M_V >$ +1.2	&	[4]	\\
No ex-companion	&	SNR 0505-67.9 in LMC	&	$M_V >$ +0.6	&	[5]	\\
No ex-companion	&	SNR 0509-68.7 in LMC	&	$M_V >$   0.0	&	[5]	\\
No ex-companion	&	SN 1006	&	$M_V >$ +4.9	&	[6]	\\
No ex-companion	&	Tycho's SN	&	$M_V >$ +5.0	&	[7]	\\
No ex-companion	&	Kepler's SN$^b$	&	$M_R >$ +3.4	&	[8]	\\
No ex-companion	&	SNR G272.2-3.2	&	$M_G >$ +7.6	&	[9]	\\
No ex-companion	&	SN 1972E in NGC 5253	&	$M_V >$ +0.5	&	[10]	\\
Ejecta/Wind impact in radio	&	SN 2011fe in M101	&	$\dot{M}_{\it wind} <$ 7.9$\times$$10^{-11}$ M$_{\odot}$ yr$^{-1}$	&	[11]	\\
Ejecta/Wind impact in radio	&	SN 2012cg in NGC 4424	&	$\dot{M}_{\it wind} <$ 5$\times$$10^{-10}$ M$_{\odot}$ yr$^{-1}$	&	[11]	\\
Ejecta/Wind impact in radio	&	SN 2014J in M82	&	$\dot{M}_{\it wind} <$ 1$\times$$10^{-10}$ M$_{\odot}$ yr$^{-1}$	&	[11]	\\
Ejecta/Wind impact in radio	&	23 normal SN{\rm I}a	&	Zero detections, $\dot{M}_{\it wind} <$ 9$\times$$10^{-8}$ M$_{\odot}$ yr$^{-1}$	&	[11]	\\
Ejecta/Wind impact in X-rays	&	SN 2020nlb in M85	&	$\dot{M}_{\it wind} <$ 9.7$\times$$10^{-10}$ M$_{\odot}$ yr$^{-1}$	&	[12]	\\
Ejecta/Wind impact in X-rays	&	SN 2017cbv in NGC 5643	&	$\dot{M}_{\it wind} <$ 7.2$\times$$10^{-10}$ M$_{\odot}$ yr$^{-1}$	&	[12]	\\
Ejecta/Wind impact in X-rays	&	53 normal SN{\rm I}a	&	Zero detections, $\dot{M}_{\it wind} <$ 1.1$\times$$10^{-6}$ M$_{\odot}$ yr$^{-1}$	&	[13]	\\
Nebular He lines	&	67 normal SN{\rm I}a	&	Zero detections, stripped mass $<$0.01 M$_{\odot}$	&	[14]	\\
		\hline
	\end{tabular}	
	
\begin{flushleft}	
\
$^a$ References: [1] Li et al. (2011), 
[2] Kelly et al. (2014), 
[3] Schaefer \& Pagnotta (2012), 
[4] Edwards, Pagnotta, \& Schaefer (2012), 
[5] Pagnotta \& Schaefer (2015), 
[6] Gonzalez-Hernandez et al. (2012), 
[7] Xue \& Schaefer (2015), 
[8] Ruiz-Lapuente et al. (2018), 
[9] Ruiz-Lapuente et al. (2023), 
[10] Do et al. (2021), 
[11] Chomiuk et al. (2016), 
[12] Sand et al. (2021), 
[13] Russell \& Immler (2012), 
[14] Tucker et al. (2020)
\\
$^b$ The shell of Kepler's SNR has some small abundance anomalies ``which can be reproduced with an asymptotic giant branch donor star with initial mass of $\sim$4 M$_{\odot}$'' (Sun \& Chen 2019).  For a second interpretation, ``the abundance ratios from the shocked ejecta are well compatible with the predicted results from spherical delayed-detonation models for Type Ia supernovae.''  And for a third interpretation, the surviving companion might be a subdwarf B star, with no giant or sub-giant companion involved.  Finally, the fourth and best interpretation is  that the Kepler SN is a core-degenerate event with no giant or sub-giant companion at the time of the explosion.  With four good explanations for the abundance anomalies, the speculation that they involve a giant companion star has no useable confidence.  Indeed, the observed limit of $M_R >$ $+$3.4 is proof already that the giant companion speculation is wrong.
 
\end{flushleft}	
	
\end{table*}

A direct method to test for helium nova progenitors is to look for the companion star long {\it after} the supernova event, when the ex-companion star can be seen free-floating near the center of the expanding supernova remnant (SNR).  In all cases, the companion will be battered by the nearby supernova explosion, but will appear at close to the same luminosity as before the explosion.  (The reason is simply that the star continues having to emit the same energy from its untouched core, so the star's base luminosity remains constant.)  The explosion site can usually be determined with usable accuracy either from the geometric center of the SNR or from the observed expansion center.  The ex-companion star was at the explosion site at the time of the eruption, and after the disappearance of the exploding star, it will be moving away from the explosion site at its orbital velocity plus a small kick.  The orbital velocity is that for a star that fills its Roche lobe, which should be around 250 km s$^{-1}$ for a companion like in V445 Pup.  The observational task is to use {\it HST} to look deep inside the central region of the SNR, seeking the ex-companion.  For the first and still-best case, SNR 0509-67.5 in the LMC has its central region empty of all stars to $V$=26.9, which forces any possible ex-companion star to be less luminous than $M_V$=$+8.4$ mag.  This extremely strict limit confidently rejects all SD models, including helium novae.  Subsequently, strict limits have been placed on any ex-companion stars for 8 other SNRs (see Table 2).  These limits are to be compared to the case for V445 Pup, with $M_V$=$-$1.9$\pm$0.7.  Further, all 9 limits exclude the possibility that the progenitor had any giant or sub-giant companion star.  So zero-out-of-nine normal SN{\rm I}a systems have any giant or sub-giant ex-companions, and this proves that helium novae cannot provide any substantial fraction of the solution to the Progenitor Problem.  Further, these nine strong limits are proof that the symbiotic star model and the recurrent nova model are not solutions to the Progenitor Problem.

A second direct method to test for helium novae is to use archival {\it HST} images from before the eruption to see whether the progenitor has a helium giant companion star.  This method has proven to be remarkably successful for measuring the nature of the progenitors for core-collapse supernovae.  For Type {\rm I}a supernovae, only the two brightest and nearest events have useful limits.  These events are SN 2011fe in M101 and SN 2014J in M82, both of which are normal SN{\rm I}a.  For these two cases, the limit on the absolute magnitude is $>$$-$1 mag.  This is to be compared to my measured $M_V$=$-$1.9$\pm$0.7 for V445 Pup.  So these two deep limits show that these two systems do not have progenitors like V445 Pup, and all possible giant stars are ruled out in general.

A good indirect method to detect a companion star is to seek the bright light created when the SN ejecta rams into the pre-eruption stellar wind of the companion.  During the eruption, the ejecta/wind interaction will be bright in the radio and in the X-ray, and this prompt radiation will be greatly brighter than anything from the supernova alone.  So the existence of a giant companion star in the progenitor will make for a prominent prompt detection of the radio and X-ray light.  And hence, the lack of any prompt radio or X-ray flux will allow for a limit to be placed on $\dot{M}_{\rm wind}$.  Table 2 collects the radio and X-ray limits on $\dot{M}_{\rm wind}$.  These quoted limits are for the fiducial and typical wind velocity of 10 km s$^{-1}$.  The five nearest normal SN{\rm I}a each individually places severe limits on the $\dot{M}_{\rm wind}$, with all being $\dot{M} <$ 9.7$\times$$10^{-10}$ M$_{\odot}$ yr$^{-1}$.  Further, general surveys have examined 23 normal SN{\rm I}a events in the radio regime and 53 normal SN{\rm I}a events in the X-ray regime, all with {\it zero} detections.  The limits are $\dot{M}_{\it wind} <$ 9$\times$$10^{-8}$ M$_{\odot}$ yr$^{-1}$ for all the radio non-detections, and $\dot{M}_{\it wind} <$ 1.1$\times$$10^{-6}$ M$_{\odot}$ yr$^{-1}$ for all the X-ray non-detections, with most supernova being far below these limits.  Allowing for overlap of individual supernovae, these studies have found strict limits on giant companion stars for 76 normal SN{\rm I}a events.  For comparison, giant stars have ordinary stellar wind rates $\dot{M}_{\rm wind}$ from 1$\times$$10^{-7}$ to 3$\times$$10^{-4}$ M$_{\odot}$ yr$^{-1}$ (Knapp \& Morris 1985), while symbiotic stars range from $10^{-8}$ to $10^{-5}$ M$_{\odot}$ yr$^{-1}$ (Seaquist \& Taylor 1990).  The result of this comparison is that any companion star in 76 normal SN{\rm I}a must have a stellar wind that is too weak to be from a giant star.  That is, the best estimate for the rate of normal SN{\rm I}a with giant companion stars is zero.  And that is zero-out-of-76, so any such progenitors must be rare, and not any solution for even a part of the Progenitor Problem.

Another strong indirect method to detect a companion star is to seek emission lines from helium during the late nebular phase of the SN, with this helium coming from the ejecta unbinding the gas off the surface of the companion, whereupon it gets entrained in the ejecta and produces emission lines.  (Similarly, this method can also seek hydrogen emission from gases stripped off the companion by the explosion, and such is sensitive to companions with ordinary solar composition.)  This entrained mass is from both ablation (heating) and stripping (momentum transfer) from the companion's surface.  The idea is to go from observed limits on the helium line flux to a limit on the mass of gas entrained into the SN ejecta.  For 73 normal SN{\rm I}a explosions, zero were seen to have any helium emission line flux (Tucker et al. 2020).  Tucker places the limit on the total mass stripped from the companion to be $<$0.023 M$_{\odot}$.  For a helium star companion with Roche lobe overflow being blasted by the supernova on the WD, Liu et al. (2013) calculates that 0.02--0.05 M$_{\odot}$ of gas is entrained into the ejecta, while Pan et al. (2012) calculates that 0.015 M$_{\odot}$ is entrained for the case of V445 Pup.  (Critically for the symbiotic and recurrent nova models, red giant companions will have $>$0.5 M$_{\odot}$ of hydrogen-rich gas entrained into the ejecta.  Such would be easily seen in all 73 normal SN{\rm I}a events, whereas zero are seen, so this constitutes a refutation of the symbiotic and recurrent nova models.)  For a conservative limit of 0.01 M$_{\odot}$, Tucker reports on 67 normal SN{\rm I}a explosions where the observational limit is better than the conservative limit for V445 Pup\footnote{Out of these, 22 are duplicates for systems already rejected by the other methods in Table 2.}.  Of these 67 supernovae, zero have any detected helium lines.  This is a strict and broad limit.  So, the best estimate for the rate of normal SN{\rm I}a with helium star companions is zero.  And that is zero-out-of-67, so any such progenitors must be rare, and not any solution for even a part of the Progenitor Problem.

I should make some comments and reasons for the three methods that are not included in Table 2:  {\bf (A)~}The ejecta/wind interaction should also produce copious amounts of ultraviolet light, above that of the supernova alone.  Brown et al. (2012, 2023) report on the lack of any detection from 29 normal  SN{\rm I}a, as viewed with the {\it Swift} XRT instrument.  Zero events were seen to be bright in the ultraviolet.  Unfortunately, the resultant limits on $\dot{M}_{\rm wind}$ are not competitive or useful.  

{\bf (B)~}A rare sub-class (labeled as `{\rm I}a-CSM', with `CSM' an abbreviation for `circumstellar medium') appears with a SN{\rm I}a spectrum plus emission lines from the ejecta ramming into a massive CSM (Silverman et al. 2013).  Such a dense CSM is not expected for any ordinary situation within a DD model, so the existence of these supernovae has been taken as evidence that at least some SN{\rm I}a are with a companion star, and hence are SD (e.g., Dilday et al. 2012).  However, there are strong evidences that the {\rm I}a-CSM events are not connected to any close-in consequential SD companion star, much less that this rare sub-class represents even a partial solution to the Progenitor Problem:  {\bf First}, roughly one-quarter of the {\rm I}a-CSM events are `late-onset' or `delayed interaction', where the ejecta takes months-to-years to impact the surrounding CSM  (Sharma et al. 2023), while traveling at $\sim$10,000 km s$^{-1}$.  At such distances from the exploding WD, the CSM is not related to any consequential companion star.  {\bf Second}, the measured mass in the CSM near the explosion site ranges from 0.4 to 5 M$_{\odot}$ (Inserra et al. 2016, Chugai \& Yungelson 2004).  Such large masses are impossible to come from any SD companion star, because the companions do not have enough mass, nor any way to create a shell of dense CSM.  {\bf Third}, two supernova models have been presented in which no SD companion star is involved.  One possibility is that the {\rm I}a-CSM events are a version of a core-collapse supernova (Inserra et al. 2016) with no companion star.  A better possibility is that the rare {\rm I}a-CSM events are examples of the core-degenerate explosion mechanism (`CD', as a distinct alternative to DD and SD), where the WD companion star was destroyed by merger at a time many years before the supernova (Soker et al. 2013, Soker 2022).  With two better alternative explanations, the {\rm I}a-CSM events are not useful evidence for the existence of companion stars in the immediate progenitor.  {\bf Fourth}, Sharma et al. (2023) measure that the {\rm I}a-CSM rate is $\sim$0.02\%--0.2\% of the overall normal SN{\rm I}a rate.  With this, {\rm I}a-CSM events are not telling us anything about the presence of companion stars for normal SN{\rm I}a, nor do they provide even a partial solution to the Progenitor Problem.  In all, {\rm I}a-CSM events cannot be used as evidence for or against companion stars in normal SN{\rm I}a progenitors.

{\bf (C)~}Kasen (2010) predicted that the presence of a companion star next to a supernova explosion could be apparent as a bright peak in the first few days of the eruption.  In particular for giant companions, for SN{\rm I}a events viewed near the line connecting the two stars, Kasen predicts a huge sudden rise at the time of the explosion, with a blue peak (a separate local maximum in the light curve) separate-from and rivaling the main SN peak, all reaching $M_V$ from $-$17 to $-$18 at a time roughly two days after the explosion.  Kasen estimates that $\sim$10\% of SN{\rm I}a with giant companions will have the viewing angle such that the Kasen-effect is prominent.  The Kasen-effect can be sought in the many wonderful full-cadence light curves that cover in time from days before the start of eruption up to the main peak.  The Kasen-effect has been sought in at least 714 normal SN{\rm I}a light curves that are easily sensitive to the blatant effects of giant companions\footnote{These include 307 SN{\rm I}a observed across the time of eruption by {\it TESS} (Fausnaugh et al. (2023), 8 observed with the {\it K2} mission (Wang et al. 2021), 3 observed with the {\it Kepler} mission (Olling et al. 2015), 127 observed with ZTF (Yao et al. 2019), 108 observed with the Sloan Digital Sky Survey (Hayden et al. 2010), 100 observed with the SuperNova Legacy Survey (Bianco et al. 2011), and 61 observed with Lick Observatory Supernova Search (Ganeshalingam et al. (2011).}.  A total of five light curves have been claimed to show the Kasen-effect (Wang et al. 2024, Dmitriadis et al. 2019a, Fausnaugh et al. 2023).  As such, the existence of the Kasen-effect might provide information on any giant companions.  {\bf First}, the best claimed Kasen-effects (for SN 2018oh and SN 2023bee) are not the predicted Kasen-peaks, nor even Kasen-bumps, but are actually barely discernible Kasen-inflections (see figures 2 and 4 top panel of Wang et al. 2024 for SN 2023bee, figure 2 of Dmitriadis et al. 2019a for SN 2018oh), while the three claimed Kasen-effects in Fausnaugh et al. (2023) are not even detectable by eye in the residual plots (see their figure 21).  Here, the important point is that the predicted Kasen-peaks for giant companions are certainly non-existent for all 714 normal SN{\rm I}a.  Kasen predicts that $\sim$71 (i.e., $\sim$10\% of 714) of these should have a prominent Kasen-peak, whereas zero are seen.  This proves that the rare and weak Kasen-inflections reported are not caused by the Kasen-effect on giant companion stars.  {\bf Second}, the five reported Kasen-inflections all have a variety of data and analysis problems that raise the question as to the existence of the Kasen-effect.  By varying the power law exponent and the time of explosion, the existence of light curve inflections can be created or eliminated (see figure 4 of Wang et al. 2024).  And simple changes in the statistical test can also create or eliminate inflections (Fausnaugh et al. 2023).  And the best fit for the claimed Kasen-inflection in SN 2023bee has a {\it reduced} chi-square of 205.5 (see table 2 of Wang et al. 2024), so either the model must be horrible or the real photometric error bars are something like 14$\times$ larger than reported, either of which makes the existence of the effect as insignificant at best.  And the {\it TESS} light curve for SN 2023bee has intermittent ``bad measurements'' that adds false flux, with one episode covering half of the claimed Kasen-inflection (see figure 11 of Wang et al. 2024), so we can have no confidence that the claimed residual-bump is not just an artifact of added flux in the decaying tail of the ``bad measurements''.  {\bf Third}, the systems with the best Kasen-effects have been proven to {\it not} have any companion star to deep limits.  For SN 2018oh and SN 2023bee, the complete absence of nebular hydrogen or helium lines rules out the possibility of any giant or sub-giant companions (Wang et al. 2024, Dimitriadis et al. 2019b).  For SN 2023bee, radio detection limits constrain the stellar wind of any progenitor, such that any giant companion is excluded (Hosseinzadeh et al. 2023).  With this, we know that the Kasen-inflections have nothing to do with giant companion stars.  {\bf Fourth}, the tiny Kasen-inflections can be well explained by an ordinary excess of $^{56}$Ni in the outer shell of the ejecta (Magee \& Maguire 2020), or by the early ejecta ramming into a CSM near the explosion (Piro \& Morozova 2016), or by the initial ignition of the thick He-shell on the surface of a sub-Chandrasekhar WD that will create radioactive material in the outer layers of ejecta (Polin et al. 2019).  With three good alternative models to explain the few small Kasen-inflections, there can be no useful confidence in using the Kasen-effect to constrain the existence and nature of companion stars of normal SN{\rm I}a.  For all four reasons, we cannot use the Kasen-inflections as evidence for the existence of giant companion stars for SN{\rm I}a progenitors.

In the end, we have five good methods to test a progenitor for giant companions, with these listed in Table 2, plus three further methods unable to make confident detections of giant companions.  After allowing for overlapping of individual supernovae in these lists, we end up with 136 separate normal SN{\rm I}a systems for which strong limits demonstrate that giant companion stars (as in V445 Pup and as in any helium nova) do not exist.  

The conclusion is that giant or sub-giant companions appear in zero-out-of-136 testable systems.  The best estimate of the rate of helium novae in progenitors is {\it zero}, although the allowed range has some small upper limit.  From Poisson statistics, the 1-sigma upper limit on the fraction of normal SNIa events is 0.84\%.  The interpretation of an exactly zero rate is that the helium nova channel is impossible, perhaps because all helium nova ejecta have very large masses like for V445 Pup.  The interpretation of a small positive rate is that the helium novae are rare.  With the 136 tested supernovae, we cannot distinguish between `impossible' and `rare'.  In all cases, helium nova constitute at most a very small fraction of normal SN{\rm I}a, and cannot comprise any part of a solution to the Progenitor Problem.

\section{THE LARGER PICTURE}

The close-up picture of V445 Pup from this paper is that of a high-mass WD in a 1.87 day orbit with a 2.65$\pm$0.35 R$_{\odot}$ giant companion that was stripped of its outer hydrogen-rich layers.  This close-up picture has the 2000 eruption making the orbital period {\it increase} by 935$\pm$27 ppm, from which we confidently know that the nova eruption ejected $\gg$0.001 M$_{\odot}$.  This $M_{\rm ejecta}$ is greatly larger than the nova trigger mass, so the V445 Pup WD must be losing mass across each eruption cycle.  With this, for a larger picture, V445 Pup cannot evolve to a supernova eruption, and is not a SN{\rm I}a progenitor.

	Now that we have a clear picture of the current state of V445 Pup, I can look at the bigger picture of its evolution as a binary star.  The basic story of V445 Pup is clear, and here is a schematic summary:  {\bf (A)~}The star formed as a wide binary with original masses of perhaps 8 and 2 M$_{\odot}$.  {\bf (B)~}When the primary star evolved off the main sequence, it expanded to engulf the secondary star and to eject a normal planetary nebula.  The common envelope phase will tighten the orbit to a smaller period.  {\bf (C)~}After the end of the common envelope, the system consisted of a WD and the original secondary still as a main sequence star.  {\bf (D)~}In the ordinary evolution, the core of the secondary ran out of hydrogen, and it expanded to form a second common envelope phase.  During this phase, the outer hydrogen-rich layers of the secondary star were stripped away, leaving exposed the helium-rich mantle.  This phase also ground down the orbital period to near 1.87 days.  {\bf (E)~}Either by ordinary angular momentum loss from the binary, or from further expansion of the helium star, the secondary star came into contact with its Roche lobe, and accretion starts.  {\bf (F)~}As accreted gas pile up on the primary star, nova eruptions repeatedly blow off huge masses away from the binary.  After each eruption, the WD is eroded, so as to be lowering M$_{\rm WD}$ over each eruption cycle.  The companion star is also losing mass each cycle due to the the dredge-up of mass during each eruption.  This is the current state of V445 Pup.  {\bf (G)~}Into the future, as time goes on, the accretion and nova-ejections will continue, with both stellar masses being whittled down in size.  {\bf (H)~}At some point, the evolution of the core of the helium star will form a CO WD and eject a planetary nebula.  {\bf (I)~}V445 Pup will end up as a WD/WD binary with a period of perhaps one day, orbiting quietly forever.
	
	While this bigger picture is clear, a variety of details are not known with any useable confidence.  The most important uncertainty is likely to be the question of the composition of the WD.  Perhaps the original primary was massive enough so that an ONe WD was formed, for which M$_{\rm WD}$ would be near the Chandrasekhar mass.  To recall, Kato et al. (2008) make a strong case that currently M$_{\rm WD}$=1.37 M$_{\odot}$ or so.  In this case, the WD we are seeing should have a surface composition with abundant neon, and then with the observed dredge-up, the eruption should be a neon nova.  For this possibility, we can only see the [Ne {\rm III}] lines by looking in the near-ultraviolet during the nebular phase long after peak, and I am aware of no spectra that covers this possibility, so V445 Pup might well be a neon nova.  The alternative is that the primary formed its CO WD with a mass $<$1.2 M$_{\odot}$ or so, and the extra mass (to get it up to 1.37 M$_{\odot}$) came from the helium star during the second common envelope stage.  With this alternative, the dredge-up will make the current ejecta have the composition of the helium star envelope, and this composition is consistent with the observed spectra during eruption.  

We can see a yet larger picture by asking whether any SN{\rm I}a systems have a companion star like that in V445 Pup.  For this question, zero out of 11 normal SN{\rm I}a systems can possibly have a companion like in V445 Pup (or any other helium nova), because no companion is visible either before or after the explosion, with severe limits that $M_V$ for any companion star is less-luminous than $-$1.9$\pm$0.7 mag.  And, zero out of 80 normal SN{\rm I}a systems can possibly have any companion like in V445 Pup (or any helium nova), because there is no detected radio or X-ray emission as required for the ejecta/wind interaction, with severe limits on the possible $\dot{M}_{\rm wind}$ of $<$ 9$\times$$10^{-8}$ M$_{\odot}$ yr$^{-1}$.  And, zero out of 67 normal SN{\rm I}a systems can possibly have any companion like in helium novae or in V445 Pup, because there is no nebular helium emission line by entrained gas from the companion, with strict limits on the stripped mass of $<$0.01 M$_{\odot}$.  In the majority of these cases, the constraints on any companion are greatly smaller than what is quoted here, often by orders-of-magnitude.  After noting duplications on these lists, we have 136 separate normal SN{\rm I}a systems that cannot have a progenitor like in V445 Pup or like in any helium nova.  This starkly shows that in all of the 136 systems where it can be tested, the helium nova case is strongly rejected.  The helium nova progenitor fraction is $<$0.84\%.  This provides a second proof that helium novae are not normal SNIa progenitors at any level that can solve the Progenitor Problem.

This larger picture can be expanded further to ask about the other SD models that require a giant or sub-giant companion.  That is, in addition to the helium nova progenitor model, the symbiotic path is a theory construct that requires a luminous and active red giant star feeding matter onto the WDs.  Further, the general recurrent nova path apparently requires 80\% or so of the progenitors to have giant or sub-giant companions, as measured by the known RN population in our Milky Way.  The observational results in Table 2 can be applied to the SD pathways involving symbiotic stars and recurrent novae, as well as for helium novae.  The conclusion is that zero out of 136 normal SN{\rm I}a systems have the giant companion required by the helium nova, symbiotic star, or recurrent nova SD models.  That is, these SD models are completely refuted as solutions to the Progenitor Problem.  It does not matter whether anyone's models suggest that these SD paths can work or not, rather, the overwhelming numbers show that these SD paths do not contribute in any recognizable numbers to the normal SNIa population seen in the sky.

	The most important aspect of the larger picture relates to the Progenitor Problem.  Two strong sets of evidence prove that V445 Pup and helium novae in general are not SN{\rm I}a progenitors, and can constitute no recognizable fraction of the solution for the Progenitor Problem.  Further, the broad and deep limits on the possibility of any giant or sub-giant companion stars rules out the symbiotic nova and the recurrent nova models.  That is, there is no chance that helium novae, symbiotic novae, or recurrent novae provide any measurable fraction of the observed normal SN{\rm I}a events.  And stepping back to see the larger picture, with the rejection of the helium nova, symbiotic nova, and recurrent nova models (the most popular and prominent SD models), the single-degenerate concept is greatly diminished as having no viable progenitors that can be pointed at.

\begin{acknowledgments}
The American Association of Variable Star Observers (AAVSO) provided a variety of useful services, including the archiving of the light curve from Monard (in the AAVSO International Database), finder charts (in VSP), accurate comparison star magnitudes (in APASS), and the discrete Fourier transform tool (in VSTAR). This research was made possible through the use of the AAVSO Photometric All- Sky Survey (APASS), funded by the Robert Martin Ayers Sciences Fund and NSF AST-1412587.  I am grateful to Berto Monard (Bronberg Observatory, Pretoria, South Africa) for his long-term light curve, both during the 2000 eruption and persistently through to recent years, with this being only a small part of his vast collection of magnitudes for interesting CVs.  I thank the observers and archivists of the HCO plate archives, and the DASCH program (J. Grindlay PI) for their huge and excellent effort at making high-quality scans of the individual plates available on-line, as well as being available for by-eye examination.  This work has made use of data provided by Digital Access to a Sky Century @ Harvard (DASCH), which has been partially supported by NSF grants AST-0407380, AST-0909073, and AST-1313370.  
\end{acknowledgments}

%

\vspace{5mm}
\facilities{DASCH, ZTF, APASS, AAVSO, TESS}


{}



\begin{thebibliography}{99}

\bibitem[\protect\citeauthoryear{Ashok and Banerjee}{2003}] {Ashok and Banerjee 2003}
Ashok, N. M., \& Banerjee, D. P. K. 2003, A\&A, 409, 1007
\bibitem[\protect\citeauthoryear{Banerjee et al.}{2023}] {Banerjee et al. 2023}
Banerjee, D. P. K., Evans, A., Woodward, C. E., et al. 2023, ApJ, 952, L26
\bibitem[\protect\citeauthoryear{Bianco et al.}{2005}] {Bianco et al. 2005} 
Bianco, F. B., Howell, D. A., Sullivan, M., et al. 2011, ApJ, 741, 20
\bibitem[\protect\citeauthoryear{Brooks et al.}{2015}] {Brooks et al. 2015}
Brooks, J., Bildsten, L., Marchant, P., \& Paxton, B. 2015, ApJ, 807, 74
\bibitem[\protect\citeauthoryear{Brown et al.}{2012}] {Brown et al. 2012} 
Brown, P. J., Dawson, K. S., Harris, D. W., et al. 2012, ApJ, 749, 18
\bibitem[\protect\citeauthoryear{Brown et al.}{2023}] {Brown et al. 2023} 
Brown, P. J., Robertson, M., Devarakonda, Y., et al. 2023, Universe, 9, 218
\bibitem[\protect\citeauthoryear{Chomiuk et al.}{2016}] {Chomiuk et al. 2016} 
Chomiuk, L., Soderberg, A. M., Chevalier, R. A., et al. 2016, ApJ, 821, 119
\bibitem[\protect\citeauthoryear{Chugai and Yungelson}{2004}] {Chugai and Yungelson 2004} 
Chugai, N. N., \& Yungelson, L. R. 2004, AstL, 30, 65
\bibitem[\protect\citeauthoryear{Dilday+}{2012}] {Dilday+ 2012}
Dilday, B., Howell, D. A., Cenko, S. B., et al. 2012, Science, 337, 942
\bibitem[\protect\citeauthoryear{Dimitriadis et al.}{2019a}] {Dimitriadis et al. 2019a} 
Dimitriadis, G., Foley, R. J., Rest, A., et al. 2019a, ApJ, 870, L1
\bibitem[\protect\citeauthoryear{Dimitriadis et al.}{2019b}] {Dimitriadis et al. 2019b} 
Dimitriadis, G., Foley, R. J., Rest, A., et al. 2019b, ApJ, 870, L14
\bibitem[\protect\citeauthoryear{Do et al.}{2021}] {Do et al. 2021} 
Do, A., Shappee, B. J., De Cuyper, J. -P., et al. 2021, MNRAS, 508, 3649
\bibitem[\protect\citeauthoryear{Edwards et al.}{2012}] {Edwards et al. 2012} 
Edwards, Z. I., Pagnotta, A., \& Schaefer, B. E. 2012, ApJ, 747, L19
\bibitem[\protect\citeauthoryear{Fausnaugh et al.}{2023}] {Fausnaugh et al. 2023} 
Fausnaugh, M. M., Vallely, P. J., Tucker, M. A., et al. 2023, ApJ, 956, 108
\bibitem[\protect\citeauthoryear{Flower}{1996}] {Flower 1996} 
Flower, P. J. 1996, ApJ, 469, 355
\bibitem[\protect\citeauthoryear{Frank, King, and Raine}{2002}]{Frank, King, and Raine 2002}
 Frank J., King A., \& Raine D. 2002, Accretion Power in Astrophysics. Cambridge
University Press, Cambridge
\bibitem[\protect\citeauthoryear{Ganeshalingam et al.}{2011}] {Ganeshalingam et al. 2011} 
Ganeshalingam, M., Li, W., \& Filippenko, A. V., 2011, MNRAS, 416, 2607
\bibitem[\protect\citeauthoryear{Goranskij et al.}{2010}] {Goranskij et al. 2010}
Goranskij, V. P., Shugarov, S. Yu., Zharova, A. V., Kroll, P., Barsukova, E. A. 2010, Peremennye Zvezdy, 30, 1
\bibitem[\protect\citeauthoryear{Hayden et al.}{2010}] {Hayden et al. 2010} 
Hayden, B. T., Garnavich, P. M., Kasen, D., et al. 2010, ApJ, 722, 1691
\bibitem[\protect\citeauthoryear{Hosseinzadeh et al.}{2023}] {Hosseinzadeh et al. 2023} 
Hosseinzadeh, G., Sand, D. J., Sarbadhicary, S. K., et al. 2023, ApJ, 953, L15
\bibitem[\protect\citeauthoryear{Iijima+}{2008}] {Iijima+ 2008}
Iijima, T. \& Nakanishi, H. 2008, A\&A, 482, 865
\bibitem[\protect\citeauthoryear{Inserra et al.}{2016}] {Inserra et al. 2016} 
Inserra, C., Fraser, M., Smartt, S. J., et al. 2016, MNRAS, 459, 2721
\bibitem[\protect\citeauthoryear{Kanatsu}{2000}] {Kanatsu 2000}
Kanatsu, K. 2000, IAUCirc, 7552
\bibitem[\protect\citeauthoryear{Kasen}{2010}] {Kasen 2010} 
Kasen, D. 2010, ApJ, 708, 1025
\bibitem[\protect\citeauthoryear{Kato and Hachisu}{2003}] {Kato and Hachisu 2003}
Kato, M. \& Hachisu, I. 2003, ApJ, 598, L107
\bibitem[\protect\citeauthoryear{Kato et al.}{2008}] {Kato et al. 2008}
Kato, M., Hachisu, I., Kiyota, S., \& Saio, H. 2008, ApJ, 684, 1366
\bibitem[\protect\citeauthoryear{Kato et al.}{1989}] {Kato et al. 1989}
Kato, M., Saio, H., \& Hachisu, I. 1989, ApJ, 340, 509
\bibitem[\protect\citeauthoryear{Kelly et al.}{2014}] {Kelly et al. 2014} 
Kelly, P. L., Fox, O. D., Filippenko, A. V., et al. 2014, ApJ, 790, 3
\bibitem[\protect\citeauthoryear{Knapp and Morris}{1985}] {Knapp and Morris 1985} 
Knapp, G. R., \& Morris, M. 1985, ApJ, 292, 640
\bibitem[\protect\citeauthoryear{Kool et al.}{2023}] {Kool et al. 2023} 
Kool, E. C., Johansson, J., Sollerman, J., et al. 2023, Natur, 617, 477
\bibitem[\protect\citeauthoryear{Li et al.}{2011}] {Li et al. 2011} 
Li, W., Bloom, J. S., Podsiadlowski, P., et al. 2011, Nature, 480, 348
\bibitem[\protect\citeauthoryear{Liu et al.}{2013}] {Liu et al. 2013}
Liu, Z.-W., Pakmor, R., Seitenzahl, I. R., et al. 2013, ApJ, 774, 37
\bibitem[\protect\citeauthoryear{Lynch et al.}{2004}] {Lynch et al. 2004}
Lynch, D. K., Rudy, R. J., Mazuk, S., \& Venturini, C. C. 2004, AJ 128, 2962
\bibitem[\protect\citeauthoryear{Magee and Macguire}{2020}] {Magee and Macguire 2020} 
Magee, M. R. \& Maguire, K. 2020, A\&A, 642, 189
\bibitem[\protect\citeauthoryear{Maoz+}{2014}] {Maoz+ 2014} 
Maoz, D., Filippo, M., \& Nelemans, G. 2014, ARA\&A, 52, 107
\bibitem[\protect\citeauthoryear{McCully et al.}{2014}] {McCully et al. 2014}
McCully, C., Jha, S. W., Foley, R. J., et al. 2014, Nature, 512, 54
\bibitem[\protect\citeauthoryear{Nieva}{2013}] {Nieva 2013}
Nieva, M. -F. 2013, A\&A, 550, 26
\bibitem[\protect\citeauthoryear{Olling et al.}{2015}] {Olling et al. 2015} 
Olling, R. P., Mushotzky, R., Shaya, E. J., et al. 2015, Natur, 521, 332
\bibitem[\protect\citeauthoryear{Pagnotta and Schaefer}{2015}] {Pagnotta and Schaefer 2015} 
Pagnotta, A., \& Schaefer, B. E. 2015, ApJ, 799, 101
\bibitem[\protect\citeauthoryear{Pan et al.}{2012}] {Pan et al. 2012} 
Pan, K.-C., Ricker, P. M., \& Taam, R. E. 2012, ApJ, 750, 151
\bibitem[\protect\citeauthoryear{Piro and Morozova}{2016}] {Piro and Morozova 2016}
Piro, A. L., \& Morozova, V. S. 2016, ApJ, 826, 96
\bibitem[\protect\citeauthoryear{Polin et al.}{2019}] {Polin et al. 2019} 
Polin, A., Nugent, P., \& Kasen, D. 2019, ApJ, 873, 84
\bibitem[\protect\citeauthoryear{Russell and Immler et al.}{2012}] {Russell and Immler et al. 2012} 
Russell, B. R., \& Immler, S. 2012, ApJ, 748, L29
\bibitem[\protect\citeauthoryear{Ruiz-Lapuente et al.}{2018}] {Ruiz-Lapuente et al. 2018} 
Ruiz-Lapuente, P., Damiani, F., Bedin, L., et al. 2018, ApJ, 862, 124
\bibitem[\protect\citeauthoryear{Ruiz-Lapuente et al.}{2023}] {Ruiz-Lapuente et al. 2023} 
Ruiz-Lapuente, P., Gonzalez Hernandez, J. I., Cartier, R., et al. 2023, ApJ, 947, 90
\bibitem[\protect\citeauthoryear{Sand et al.}{2021}] {Sand et al. 2021} 
Sand, D. J., Sarbadhicary, S. K., Pellegrino, C., et al. 2021, ApJ, 922, 21
\bibitem[\protect\citeauthoryear{Schaefer}{2011}] {Schaefer 2011} 
Schaefer, B. E.  2011, ApJ, 742, 113     
\bibitem[\protect\citeauthoryear{Schaefer}{2020}] {Schaefer 2020} 
Schaefer, B. E. 2020, MNRAS, 492, 3323		
\bibitem[\protect\citeauthoryear{Schaefer}{2022}] {Schaefer 2022} 
Schaefer, B. E. 2022, MNRAS, 517, 6150		
\bibitem[\protect\citeauthoryear{Schaefer}{2023}] {Schaefer 2023} 
Schaefer, B. E. 2023, MNRAS, 525, 785		
\bibitem[\protect\citeauthoryear{Schaefer}{2024}] {Schaefer 2024} 
Schaefer, B. E. 2024, ApJ, 966, 155		
\bibitem[\protect\citeauthoryear{Schaefer and Pagnotta}{2012}] {Schaefer and Pagnotta 2012} 
Schaefer, B. E. \& Pagnotta, A. 2012, Nature, 481, 164
\bibitem[\protect\citeauthoryear{Seaquist and Taylor}{1990}] {Seaquist and Taylor 1990} 
Seaquist, E. R., \& Taylor, A. R. 1990, ApJ, 349, 313
\bibitem[\protect\citeauthoryear{Sharma et al.}{2023}] {Sharma et al. 2023} 
Sharma, Y., Sollerman, J., Fremling, C., et al. 2023, ApJ, 948, 52
\bibitem[\protect\citeauthoryear{Shen and Bildsten}{2009}] {Shen and Bildsten 2009} 
Shen, K. J., \& Bildsten, L. 2009, ApJ, 699, 1365
\bibitem[\protect\citeauthoryear{Shen and Quataert}{2022}] {Shen and Quataert 2022} 
Shen, K. J., \& Quataert, E. 2022, ApJ, 938, 31
\bibitem[\protect\citeauthoryear{Shimamoto et al.}{2017}] {Shimamoto et al. 2017} 
Shimamoto, S., Sakon, I., Onaka, T., et al. 2017, Publ. Korean Astr. Soc., 32, 109
\bibitem[\protect\citeauthoryear{Silverman et al.}{2013}] {Silverman et al. 2013} 
Silverman, J. M., Nugent, P. E., Gal-Yam, A., et al. 2013, ApJS, 207, 3
\bibitem[\protect\citeauthoryear{Soker}{2022}] {Soker 2022} 
Soker, N. 2022, RAA, 22, 035025
\bibitem[\protect\citeauthoryear{Soker et al.}{2013}] {Soker et al. 2013} 
Soker, N., Kashi, A., Garcia-Berro, E., Torres, S., \& Camacho, J. 2013, MNRAS, 431, 1541
\bibitem[\protect\citeauthoryear{Sun and Chen}{2019}] {Sun and Chen 2019} 
Sun, L., \& Chen, Y. 2019, ApJ, 872, 45
\bibitem[\protect\citeauthoryear{Tucker et al.}{2020}] {Tucker et al. 2020} 
Tucker,  M. A., Shappee, B. J., Vallely, P. J., et al. 2020, MNRAS, 493, 1044
\bibitem[\protect\citeauthoryear{Wang et al.}{2024}] {Wang et al. 2024} 
Wang, Q., Rest, A., Dimitriadis G., et al. 2024, ApJ, 962, 17
\bibitem[\protect\citeauthoryear{Wang et al.}{2021}] {Wang et al. 2021} 
Wang, Q., Rest, A., Zenati, Y., et al. 2021, ApJ, 923, 167
\bibitem[\protect\citeauthoryear{Woudt et al.}{2009}] {Woudt et al. 2009} 
Woudt, P. A., Steeghs, D., Karovska, M., et al. 2009, ApJ, 706, 738
\bibitem[\protect\citeauthoryear{Xue and Schaefer}{2015}] {Xue and Schaefer 2015} 
Xue, Z., \& Schaefer, B. E. 2015, ApJ, 809, 183
\bibitem[\protect\citeauthoryear{Yao et al.}{2019}] {Yao et al. 2019} 
Yao, Y., Miller, A. A., Kulkarni, S. R., et al. 2019, ApJ, 886, 152
\bibitem[\protect\citeauthoryear{Yaron et al.}{2005}] {Yaron et al. 2005} 
Yaron, O., Prialnik, D., Shara, M. M., \& Kovetz, A. 2005, ApJ, 623, 398



\end{thebibliography}
\end{document}